\documentclass[prb,showpacs,twocolumn,aps,superscriptaddress,a4paper]{revtex4-1}
\usepackage{dcolumn,amssymb,amsmath,amsfonts,graphicx,latexsym,color,braket}

\begin{document}

\title{Zeros of Loschmidt echo in the presence of Anderson localization}
\author{Honghao Yin}
\affiliation{Department of Physics, Zhejiang Normal University, Jinhua 321004, People's Republic of China}
\author{Shu Chen}
\affiliation{Beijing National Laboratory for Condensed Matter Physics,
Institute of Physics, Chinese Academy of Sciences, Beijing 100190, China}
\author{Gao Xianlong}
\affiliation{Department of Physics, Zhejiang Normal University, Jinhua 321004, People's Republic of China}
\author{Pei Wang}
\email{wangpei@zjnu.cn}
\affiliation{Department of Physics, Zhejiang Normal University, Jinhua 321004, People's Republic of China}
\date{\today}

\begin{abstract}
We study the Loschmidt echo and the dynamical free energy of
the Anderson model after a quench of the disorder strength.
If the initial state is extended and the eigenstates of the post-quench Hamiltonian are strongly localized,
we argue that the Loschmidt echo exhibits zeros periodically with the period $2\pi /D$ where
$D$ is the width of spectra.
At these zeros, the dynamical free energy diverges in a logarithmic way.
We present numerical evidence of our argument in one- and three-dimensional Anderson models.
Our findings connect the dynamical quantum phase transitions
to the localization-delocalization phase transitions.
\end{abstract}

\maketitle

\section{\label{sec:level1}Introduction}

Since Anderson's seminal paper in 1958~\cite{anderson},
Anderson localization has been extensively studied. In recent years, great progress was made
in simulating the nonequilibrium dynamics of closed quantum systems
by using ultracold atoms~\cite{bloch}. A question then arises as to what is the influence of
Anderson localization on the nonequilibrium
dynamics. An especially interesting protocol of
driving a system out of equilibrium is by a quantum quench,
i.e. by suddenly changing the Hamiltonian of the system.
For a homogeneous integrable system such as a noninteracting Fermi gas,
the local observable relaxes to a steady value after a quench~\cite{rigol06,rigol07}.
And this steady value can be predicted by the generalized Gibbs ensemble (GGE)~\cite{rigol06}.
But if the post-quench Hamiltonian has localized eigenstates,
the observable exhibits an everlasting oscillation with its average
deviating significantly from the prediction of GGE~\cite{caneva11,gramsch,ziraldo,
ziraldo13,he}. This everlasting
oscillation comes from the pure-point spectrum associated to
the localized eigenstates~\cite{ziraldo}.

An important quantity characterizing the nonequilibrium dynamics
is the Loschmidt echo
\begin{equation}
 \mathcal{L}(t)= \bra{\Psi(0)}e^{-i\hat H t} \ket{\Psi(0)},
\end{equation}
where $\ket{\Psi(0)}$ denotes the pre-quench quantum state and $\hat H$
the post-quench Hamiltonian.
$\mathcal{L}(t)$ might become zero at some critical times $t^*$, at which
the dynamical free energy in thermodynamic limit is nonanalytic.
The dynamical free energy is usually defined as
\begin{equation}\label{eq:defft}
 f(t) = - \lim_{N\to \infty} \frac{1}{N} \ln \left|\mathcal{L} (t)\right|^2,
\end{equation}
where $N$ denotes the number of particles. The nonanalyticity of $f(t)$ at $t=t^*$
is dubbed a dynamical quantum phase transition (DQPT)~\cite{heyl}.

Ever since the original paper of Heyl {\it et al.}~\cite{heyl}, the DQPTs have been
theoretically addressed in various models~\cite{CK,fagotti,Eck,FA,Jpg,MSc,HeylT,dora,heyl14} with the experimental observations
also realized recently~\cite{flaschner,jurcevic}.
It was argued that the Loschmidt echo
exhibits zeros if the pre-quench and post-quench
Hamiltonians have topologically-different ground states~\cite{dora}, or if
the pre-quench state breaks some symmetry which is recovered after the quench~\cite{heyl14}.
The relation between the DQPTs and the topological phase transitions (or the symmetry-breaking phase transitions)
is then established.

While most theoretical works on DQPTs focus on the quantum systems with
phase transitions caused by broken symmetries or accompanied by the
close of energy gap, DQPTs driven by disorders are rarely studied except of a recent work on the one-dimensional
Aubry-Andr\'{e} model with quasi-periodic potentials~\cite{yang}.
In this paper we make a further step towards connecting the DQPTs to the
localization-delocalization transitions. This is done by
proving that zeros of Loschmidt echo and nonanalyticity
of the dynamical free energy occur if an extended initial state is quenched into
a strongly-localized regime. Besides a proof based on the properties
of the wave functions and the spectrum in a localized phase,
we also provide numerical evidence in one- and three-dimensional Anderson
models. Different from the quasi-periodic system with deterministic quasi-random potentials, we will discuss
the more general disordered systems with random potentials of white-noise type
and focus our study on the quench dynamics from an initial
extended state to the strongly-disordered regime. We find that
the critical times in the case of white-noise potentials are $t^*_n=\frac{2\pi n}{D}$
with $n$ an integer and $D$ the spectrum width of $\hat H$, which
are different from the critical times in the case of
quasi-periodic potentials - the zeros of the Bessel function~\cite{yang}.
Our results will serve as a benchmark for understanding
the characteristics of the Loschmidt echo and the dynamical free energy after a more general quench
with the initial and the post-quench Hamiltonians at arbitrary disorder strength.

The contents of the paper are arranged as follows. In Sec.~\ref{sec:analytic}, we
present a general argument about the zeros of the Loschmidt echo
and the nonanalyticity of the dynamical free energy. The numerical evidence
is given in Sec.~\ref{sec:level3}. Sec.~\ref{sec:conclusion} is a short summary.

\section{\label{sec:analytic} DQPT in the strong-disorder limit}

The Hamiltonian of the Anderson model is in general expressed as
\begin{equation}\label{eq:ham}
\hat{H}=-g\sum_{\langle i,j \rangle}(\hat{a}^{\dag}_{i}\hat{a}^{}_{j}+H.c.)
+\sum_{i}u_{i}\hat{a}^{\dag}_{i}\hat{a}^{}_{i},
\end{equation}
where $\langle i,j\rangle$ denotes a pair of neighbor sites and
$u_{i}$ is the on-site potential which is an independent random number
distributed uniformly in the interval $[-W/2, W/2]$ with $W$ denoting
the disorder strength. $g$ is set to the unit of energy throughout the paper.
We focus on bosons in this paper. $\hat{a}^{\dag}_{i}$ and $\hat{a}_{i}$
denote the bosonic creation and annihilation operators, respectively.

Let us suppose that the system is initially prepared in an extended
state, e.g., in the ground state of the Hamiltonian~(\ref{eq:ham}) at $W=0$.
In the initial state, all the bosons occupy the lowest
energy level, the single-particle wave function of which is denoted as $\ket{\phi(0)}$.
The system is then driven out of equilibrium by suddenly changing the Hamiltonian
to $\hat H$ at finite $W$. Since we do not consider the interaction between bosons,
the many-body wave function keeps a product state during the time evolution. Furthermore,
all the bosons have the same wave function at arbitrary time, which is denoted as $ \ket{\phi(t)}$.
We use $\ket{\alpha_n}$ to denote the single-particle eigenstate of the post-quench Hamiltonian with the corresponding
eigenenergy $\epsilon_n$, where $n=1,2,\cdots,L$ and $L$ is the total number of sites. We then obtain
\begin{equation}
 \ket{\phi(t)} = \sum_{n=1}^L \braket{\alpha_n | \phi(0)} e^{-i\epsilon_n t}
 \ket{\alpha_n}.
\end{equation}
If the system contains $N$ bosons, the Loschmidt echo of the many-body wave function becomes
\begin{equation}
\begin{split}
 \mathcal{L}(t) = & \braket{\phi(0)|\phi(t)}^N \\
 = &  \left( \sum_{n=1}^L \left| \braket{\phi(0)|\alpha_n}\right|^2
 e^{-i\epsilon_n t} \right)^N.
 \end{split}
\end{equation}

Now we discuss the case in which the system is quenched into the strongly-disordered regime.
In this regime, all the single-particle eigenstates $\ket{\alpha_n}$
are strongly-localized like a $\delta$-function. Recall that the initial
single-particle state $\ket{\phi(0)}$ is extended over the whole system like
a plane wave. Therefore, the overlap $\braket{\phi(0)|\alpha_n}$ is approximately $1/\sqrt{L}$.
With this in mind, we find that $\left| \braket{\phi(0)|\alpha_n}\right|^2$ is $1/L$
which is independent of $n$. And the Loschmidt echo becomes $\mathcal{L}(t)= l(t)^N$,
where the single-particle Loschmidt echo is
\begin{equation}
 {l}(t)=\frac{1}{L} \sum_{n=1}^L e^{-i\epsilon_n t}.
\end{equation}

The sum of $e^{-i\epsilon_n t}$ is determined by the single-particle levels $\epsilon_n$,
which depend on the configuration of disorder and are in fact random numbers. Let us study
the joint probability density $P(\epsilon_1,\epsilon_2,\cdots,\epsilon_L)$
of these random levels. In the Hamiltonian~(\ref{eq:ham}),
the disordered potential $u_i$ is a random number uniformly
distributed in the range from $-W/2$ to $W/2$. In the localized phase,
the probability density of the nearest-neighbor level spacing is the
Poisson distribution ($\sim e^{-s}$) which results from
$\epsilon_1, \cdots, \epsilon_L$ being independent and uniformly-distributed random
numbers~\cite{altshuler,izrailev,shklovskii}.
Therefore, in the localized phase we have
\begin{equation}
\begin{split}\label{eq:dos}
 P(\epsilon_1,\epsilon_2,\cdots,\epsilon_L) = 1/D^L ,
 \end{split}
\end{equation}
where $D$ is the width of the single-particle spectra. In the strong disorder
limit (large $W$ limit), the disordered potentials govern the Hamiltonian~(\ref{eq:ham}),
and $\epsilon_1,\cdots,\epsilon_L$ are then no more than the disordered potentials which are of course
independent and uniformly-distributed random numbers according to the definition.
In this limit the width of spectra $D$ is equal to $W$.

Equipped with the knowledge of $P(\epsilon_1,\epsilon_2,\cdots,\epsilon_L)$,
we can now calculate the Loschmidt echo, which after averaged over the level distribution
becomes
\begin{equation}
\begin{split}\label{eq:analosch}
 {l}(t)=& \int_{-D/2}^{D/2} d\epsilon_1 \cdots d\epsilon_L
 P(\epsilon_1,\epsilon_2,\cdots,\epsilon_L)\frac{1}{L} \sum_{n=1}^L e^{-i\epsilon_n t}\\
 = & \frac{e^{-itD/2}\left(e^{itD}-1\right)}{iDt}.
 \end{split}
\end{equation}
It is clear that both the single-particle and many-body Loschmidt echo
vanish periodically at the critical times $t^*_n = 2\pi n/D$ with $n=1,2,\cdots$
a positive integer. The vanish of Loschmidt echo causes the nonanalyticity
of the dynamical free energy which according to Eq.~(\ref{eq:defft}) evaluates
\begin{equation}\label{eq:dfe}
f(t) = -\ln \frac{\sin^2\left(Dt/2\right)}{\left(Dt/2\right)^2}.
\end{equation}
The dynamical free energy is divergent at the critical times $t^*_n$.
Both the Loschmidt echo and the dynamical free energy signal
periodically-occurred DQPTs at $t^*_n = 2\pi n/D$.
And the dynamical free energy is not continuous at these DQPTs.

\section{\label{sec:level3}DQPTs in one- and three-dimensional Anderson models}

We numerically study the Loschmidt echo and the dynamical free energy
for the one- and three-dimensional Anderson Hamiltonians.
In our calculation we choose the periodic boundary condition.
At $W=0$, the single-particle eigenstates of the Anderson model are plane waves.
In one dimension, the eigenstates are all localized in the presence of infinitesimal $W$.
While in three dimensions, the eigenstates are localized only if the disorder strength $W$
is beyond a critical value $W_c$. The initial state is set to the ground state of $\hat H$
at $W=0$. A finite $W$ is switched on at $t=0$ and the Loschmidt echo
and the dynamical free energy are then calculated.
In the calculation, we perform an average over different disorder configurations by
sampling the disorder potentials for many times until the results converge.

\begin{figure}[h]
  \centering
  \includegraphics[width=0.55\textwidth]{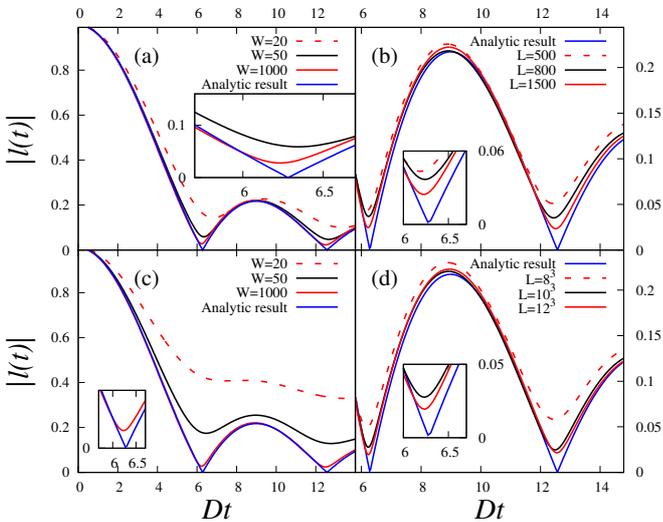}
 \caption{The absolute value of the Loschmidt echo as a function of time.
 The panels (a) and (b) are for the one-dimensional Anderson model
  at different $W$ with fixed $L=1000$ and at different $L$ with fixed $W=1000$, respectively.
  The panels (c) and (d) are for the three-dimensional Anderson
  model at different $W$ with fixed $L=1000$ and at different $L$ with fixed $W=1000$, respectively.
  The analytic result, i.e. Eq.~(\ref{eq:analosch}), is also plotted for comparison.
  In the vicinity of $Dt=2\pi, 4\pi$, we see
  that $|l(t)|$ approaches zero with increasing $L$ or $W$.}\label{1}
\end{figure}
Let us first see the single-particle Loschmidt echo $l(t)=\braket{\phi(0)|\phi(t)}$.
The single-particle eigenenergy at $W=0$ is $\epsilon = -2g
\cos k$ in one dimension or $\epsilon= -2g \left( \cos k_x + \cos k_y + \cos k_z\right)$ in three dimensions,
thereafter, the ground state $\ket{\phi(0)}$ is a plane wave of zero wave vector. The amplitude of $\ket{\phi(0)}$ at arbitrary site
is $1/\sqrt{L}$. Figs.~\ref{1}(a) and~(b) display
the absolute value $\left| l(t)\right|$ for the one-dimensional Anderson model
at different $W$ and $L$.
The numerical results fit well with Eq.~(\ref{eq:analosch})
if the disorder strength is strong ($W>50$). For fixed $W$, the
fit between numerics and Eq.~(\ref{eq:analosch}) becomes even better
as the system's size increases. As $L$ increases, the minimum of $l(t)$
goes towards zero. We then expect that in the limit $L\to\infty$ the Loschmidt
echo does become zero at some critical times. And the value of the critical times
also fits well with our prediction, i.e., $t^*_n D = 2\pi n$.

Fig.~\ref{1}(c) and~(d) show the absolute value of the Loschmidt echo for the three-dimensional
Anderson model at different $W$ and $L$.
In both one-dimensional and three-dimensional cases, the
Loschmidt echo always fits Eq.~(\ref{eq:analosch}) at large $W$, e.g. at $W=1000$.
By analyzing the change of $l(t)$ with increasing $L$,
we also find in three dimensions that in the limit $L\to \infty$
the single-particle Loschmidt echo becomes zero periodically at
the critical times $t^*_n =2\pi n/D$.

Next we discuss the dynamical free energy whose nonanalyticity
unambiguously defines the DQPTs and the critical times. The dynamical free energy is related to
the single-particle Loschmidt echo by $f(t)=-\left\langle \ln \left( \left|l(t)\right|^2\right) \right\rangle$
where $\langle\rangle$ denotes the average over different disorder configurations.

\begin{figure}[h]
  \centering
  \includegraphics[width=0.5\textwidth]{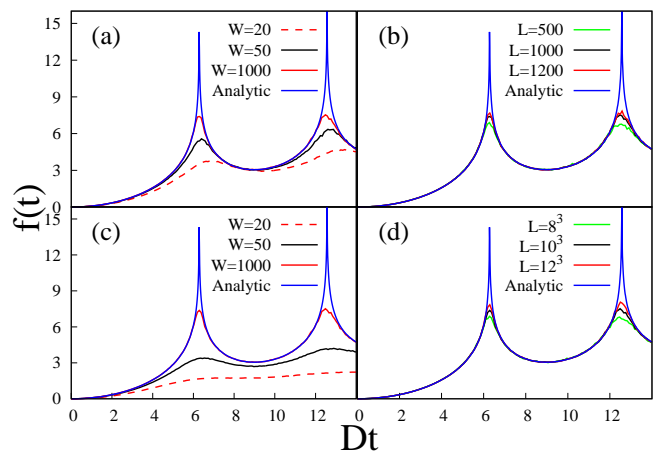}
 \caption{The panels (a) and (b) plot the dynamical free energy $f(t)$ for the one-dimensional Anderson model
 at different $W$ with fixed $L = 1000$ and at different $L$ with fixed $W = 1000$, respectively.
 The panels (c) and (d) plot $f(t)$ for the three-dimensional Anderson model at different $W$ with fixed $L=1000$
 and at different $L$ with fixed $W=1000$, respectively.}\label{2}
\end{figure}

Fig.~\ref{2} shows the dynamical free energy at different $L$ and $W$. Eq.~(\ref{eq:dfe})
is plotted at the same time for comparison, in which $D$ is also averaged over different disorder configurations.
The numerical results fit well with Eq.~(\ref{eq:dfe}) at large $W$, e.g. at $W=1000$ (see Fig.~\ref{2}(a) and~(c)).
As the disorder strength increases, the fit becomes even better.
It is clear that the dynamical free energy displays a peak periodically at the critical times $t^*_n D=2\pi n$.
We also compare the dynamical free energy at different $L$.
As the system's size increases, the peak of $f(t)$ becomes higher,
and the shape of $f(t)$ is closer to that of Eq.~(\ref{eq:dfe}) (see Fig.~\ref{2}(b) and~(d)).

The similar nonanalytic behavior of $f(t)$ has been observed in many other models, such
as the XXZ model~\cite{heyl14}. Here we would like to
emphasize that the nonanalyticity of $f(t)$ in the Anderson models is
caused by the localization-delocalization transition, different from the DQPTs
caused by broken symmetries in the XXZ model~\cite{heyl14}
or by the close of the energy gap in the topological insulators\cite{dora}.

\begin{figure}[h]
  \centering
  \includegraphics[width=0.48\textwidth]{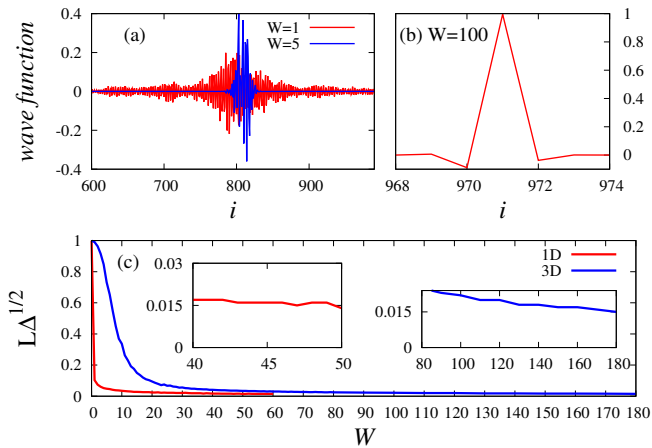}
 \caption{The panels (a) and (b) plot the $618$-th eigen wave functions
(count from the bottom of the spectrum) at different $W$ with fixed $L = 1000$
for the one-dimensional Anderson model. The panel (c) plots $L\Delta^{1/2}$ as a function
of $W$ for the one- (red) and three-dimensional (blue) Anderson models.}\label{3}
\end{figure}
At small $W$ (e.g., at $W=20$ for one dimension or
at $W=50$ for three dimensions), we clearly see the difference between numerics and Eq.~(\ref{eq:analosch})
or Eq.~(\ref{eq:dfe}) (see Fig.~\ref{1}(a),(c) and Fig.~\ref{2}(a),(c)). This difference is caused by the failure of the two
assumptions in the above derivation of Eq.~(\ref{eq:analosch}) and~(\ref{eq:dfe}). First, the
single-particle eigenstates are not $\delta$-like at small $W$ so that
their overlap with the initial wave function ($\left|\braket{\phi(0)|\alpha_n}\right|^2$)
deviates from $1/L$ but depends on $\ket{\alpha_n}$.
Fig.~\ref{3}(a) and~(b) plot the eigen wave functions of the one-dimensional Anderson Hamiltonian
at different $W$. At $W=1$ and $W=5$, the eigen wave function deviates obviously
from the $\delta$-function, even if it is approximately the $\delta$-function at $W=100$.
We quantify the deviation of $\left|\braket{\phi(0)|\alpha_n}\right|^2$ from $1/L$
by defining the overlap variance
\begin{equation}
 \Delta = \frac{1}{L^{2}}\sum_{n=1}^L \left( \left|\braket{\phi(0)|\alpha_n}\right|^2-\frac{1}{L}\right)^2.
\end{equation}
$\Delta$ is zero if $\left|\braket{\phi(0)|\alpha_n}\right|^2$ equals $1/L$ independent of $\ket{\alpha_n}$,
but $\Delta$ is finite if $\left|\braket{\phi(0)|\alpha_n}\right|^2$ deviates from $1/L$.
In Fig.~\ref{3}(c), we see $\Delta = 1/L^2$ at $W = 0$ and decays towards zero
as $W$ increases. In one dimension, the overlap variance already decays to $0.015$
at $W=50$. But the decay of $\Delta$ is much slower in three dimensions, and
it is at $W=180$ when $\Delta$ decays to $0.015$.
This explains why the deviation of the numerics from the analytical results at small $W$
is much larger in three dimensions than that in one dimension.

\begin{figure}[h]
  \centering
  \includegraphics[width=0.48\textwidth]{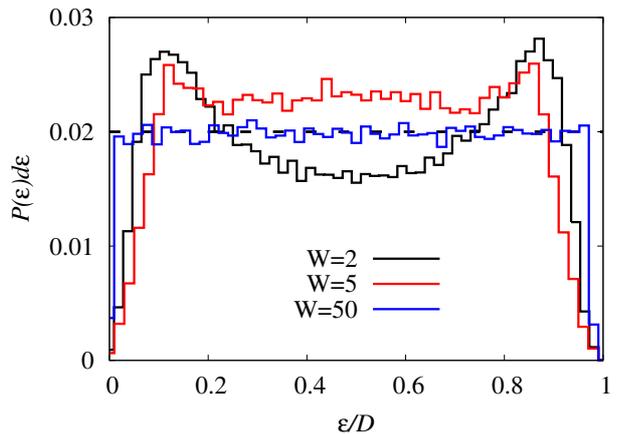}
 \caption{$P(\epsilon)$ at different disorder strength.
 We select $d\epsilon = 0.02D$ and count the number of states
 in the interval $[\epsilon , \epsilon+d\epsilon]$.
 The dashed line represents the value of $P(\epsilon)d\epsilon$ as $P(\epsilon)=1$ is a constant.}\label{4}
\end{figure}
Second, $P(\epsilon_1,\epsilon_2,\cdots,\epsilon_L)$
in Eq.~(\ref{eq:dos}) is not a constant at small $W$. Since $P(\epsilon_1,\epsilon_2,\cdots,\epsilon_L)$
is a multivariate function which is difficult to plot, we alternatively plot $P(\epsilon)=\int d\epsilon_2 \cdots d\epsilon_L
P(\epsilon,\epsilon_2,\cdots,\epsilon_L)$ in Fig.~\ref{4} which is no more than the
probability density of single-particle levels (or the normalized density of states).
If $P(\epsilon_1,\epsilon_2,\cdots,\epsilon_L)$ is a constant, $P(\epsilon)$ must also be a constant.
We see that $P(\epsilon)$ does be approximately a constant as $W$ is as large as $50$,
but apparently varies in the spectrum with two peaks at the edge and a valley at the center
for small $W$ (e.g. $W=2$).
This is not difficult to understand. At small $W$, the
hopping term in the Hamiltonian~(\ref{eq:ham}) dominates. It is well known that
the density of states has two peaks at the edge of the spectrum for the hopping Hamiltonian $\hat H_{hop}
= -g \sum_{\langle i,j\rangle} \left(\hat a^\dag_i \hat a_j + H.c.\right) $. But at large $W$, the random potentials
in the Hamiltonian~(\ref{eq:ham}) dominate, leading to a constant
density of states, as we argued below Eq.~(\ref{eq:dos}).

\section{\label{sec:conclusion}Conclusion and outlook}

In this paper, we study the Loschmidt echo and the dynamical
free energy after a quench of white-noise potentials.
We find the periodically-occurred dynamical quantum phase transitions with a period $2\pi/D$
where $D$ is the width of the spectrum.
By using the properties of the wave functions and the spectra in the strongly-localized region,
we argue that the dynamical free energy is divergent at DQPTs.
We present numerical results
in one- and three-dimensional Anderson models to support our arguments.

Finally, we would like to mention that the measurement
of the absolute value of the Loschmidt echo $\left|\mathcal{L}(t)\right|$
has been experimentally realized~\cite{flaschner,jurcevic}. In a string
of ions simulating interacting transverse-field Ising models by using
$\text{Ca}^+$ ions, the non-equilibrium dynamics at DQPTs induced by a
quantum quench is detected directly by measuring $\left|\mathcal{L}(t)\right|$
while projecting the many-body wave function at arbitrary time onto a chosen initial state\cite{jurcevic}.
With time-resolved state tomography, the topological DQPTs
of ultracold atoms in optical lattices is measured by a full access
to the evolution of the wave function~\cite{flaschner}. In view of the experimental
realization of the Anderson model in the optical lattice~\cite{BILLY,ROATI}, we then expect
that a new type of the nonanalyticity of the dynamical
free energy caused by the localization-delocalization transition predicted in this paper can be observed in cold atoms.

\section*{acknowledgments}
This work is supported by NSF of China under Grant Nos. 11304280, 11774315 and 11374266,
and the Program for New Century Excellent Talents in University. S. C. is supported by the National Key Research and Development Program of China (2016YFA0300600) and NSFC under Grants No. 11425419, No. 11374354 and No. 11174360.

\end{document}